\documentclass[prb, showpacs, twocolumn, floatfix,superscriptaddress]{revtex4}
\usepackage{bm, graphicx, amsmath}

\begin{document}

\title{Semiclassical dynamics of domain walls

in the one-dimensional Ising ferromagnet in a transverse field}

\author{B.~A.~Ivanov}
\affiliation{Institute of Magnetism National Academy of Sciences and
Ministry of Education and Science of Ukraine, 36-B Vernadskii
avenue, 03142 Kiev, Ukraine}

\author{H.-J. Mikeska}
\affiliation{Institut f\"ur Theoretische Physik, Universit\"at
Hannover, Appelstra{\ss}e 2, 30167 Hannover, Germany}

\date{\today}

\begin{abstract}
We investigate analytically and numerically the dynamics of domain
walls in a spin chain with ferromagnetic Ising interaction and subject
to an external magnetic field perpendicular to the easy magnetization
axis (transverse field Ising model). The analytical results obtained
within the continuum approximation and numerical simulations performed
for discrete classical model are used to analyze the quantum
properties of domain walls using the semiclassical approximation. We
show that the domain wall spectrum shows a band structure consisting
of 2$S$ non-intersecting zones.
\end{abstract}

\maketitle

\section{Introduction} \label{s:introduction}

Nonlinear topologically nontrivial excitations, or solitons, are known
to exist in lower-dimensional magnetic systems. There is both
theoretical and experimental evidence of solitons and in some cases
soliton effects dominate the thermodynamic behavior of one-dimensional
(1D) magnets (spin chains).  Soliton effects possibly emerged first
when absence of long range order in lower-dimensional magnets was
proved: From simple entropy arguments it has been shown that long
range magnetic order is not possible in one-dimensional magnets at
nonzero temperature and it became clear later that kink-type solitons,
which in fact are magnetic domain walls (DW), must be considered as
elementary excitations at nonzero temperature in one-dimensional
magnets, for reviews see \cite{mikeska,sowsci}.  Whereas solitons in
one-dimensional magnets have not been directly observed, dynamic
soliton effects such as soliton motion and the soliton-magnon
interaction result in soliton contributions to the dynamic response
functions, which can be studied experimentally. For example, solitons
contribute to the specific heat and to the linewidth of electron spin
resonance and the translational motion of kinks leads to the so-called
soliton central peak, which can be detected through neutron scattering
experiments.\cite{mikeska,sowsci}

In most approaches solitons have been considered using classical
continuum models, such as the Landau-Lifshitz equations or the sine
Gordon equation, see Refs.~\onlinecite{mikeska,sowsci,phys.rep}. On
the other hand, in order to describe a material such as CsCoCl$_3$ the
XXZ-model with spin $S = 1/2$ has to be used, and solitons occur as
quantum objects.\cite{villain,mikeska+} Even for the material
CsNiF$_3$, the well-known standard example for classical solitons in
1D magnets, quantum effects due to its spin $S=1$ are
essential.\cite{mikeska} Although a detailed analysis concentrated on
classical models, considerable achievements in the field of quantum
solitons were obtained rather early: We note, first of all, that the
first nonlinear excitations (spin complexes) were investigated by
Bethe in the isotropic one-dimensional ferromagnet as long ago as
1931, essentially in parallel with the prediction of
magnons.\cite{bethe} Currently it is established that the more general
XYZ-model with spin $S=1/2$ is exactly integrable, and the quantum
nonlinear excitations (spin complexes) are known from the solution of
this model.\cite{baxter}

At first sight quantum and classical solitons constitute essentially
different objects. The main property of classical solitons (or more
precisely: solitary waves) is localization, while a quantum soliton is
characterized by a definite value of the quasi-momentum $P$ and in
virtue of this it is spatially delocalized. However, this
contradiction is removed if one investigates the spin deviation
localization in the coordinate system with the origin moving with the
group velocity of the soliton.\cite{phys.rep} The comparison of
classical and quantum solitons\cite{phys.rep} reveals a striking
feature of the XYZ-model: the dispersion law (the dependence of energy
$E$ on the momentum $P$) of a spin complex in this model with spin
$S=1/2$ exactly coincides with the corresponding dependence found for
the soliton in the classical Landau-Lifshitz equations. It is evident
that such an exact correspondence cannot be a general rule; probably
it is associated with the exact integrability of both models. On the
other hand, it has become clear that if one compares the
characteristics which are relevant for both quantum and classical
approaches, first of all the dispersion law $E = E(P)$, then there are
no fundamental distinctions between quantum and classical
solitons. Renewed interest in the problem of quantum properties of
domain walls was stimulated due to the study of quantum tunneling DW
chirality effects,\cite{ivkol,br-loss,TakagiTat96,TakagiShib00,Freire}
and also by the prediction of new effects of destructive interference
in kink tunneling between neighboring crystal lattice
sites\cite{br-loss} and Bloch oscillations of the solitons.\cite
{KirLoss98}

The structure and the properties of solitons in the one-dimensional
Ising model in the presence of a transverse magnetic field have been
studied some time ago.\cite{mik-miyash} It has been demonstrated in
the semiclassical limit of large spin values $S \gg 1$ how classical
localization correlates with quantum delocalization. Within the
framework of the quantum approach the emergence of a band structure
has been revealed. However, the quantum kink dispersion law found in
this work has not been compared with the classical one. In
Ref.~\onlinecite{mik-miyash} only static solutions of the classical
problem have been used, and the comparison of the quantum kink
dispersion law with those obtained within the semiclassical
quantization model has not been made.

In this article we investigate analytically and numerically the
dynamics of domain walls in a spin chain with ferromagnetic Ising
interaction and subject to a magnetic field perpendicular to the easy
magnetization axis (transverse field Ising model). Analytical results
obtained within the continuum approximation and numerical simulations
performed for the discrete classical model are used for the analysis
of the semiclassical dynamics of domain walls including an account of
the lattice pinning potential. We show that the spectrum of the DW in
the classical continuum approximation is characterized by a periodic
dependence of the kink energy on its momentum. This produces a number
of non-trivial features for the DW motion in the presence of the
pinning potential. The quantum properties of the domain wall are
discussed on the basis of the semiclassical approximation. For the
semiclassical dynamics the role of lattice pinning effects in the
formation of the band structure of the domain wall spectrum is
investigated. The presence of the periodic character of the DW
dispersion law with lattice pinning taken into account results in a
band structure for $E(P)$ with $2S$ non-intersecting branches.

The outline of this article is as follows. In
Sec.~\ref{s:discrete} we formulate the discrete classical model
and study the DW properties within this model. The main results of
this section are the introduction of the DW coordinate, $X$, and
the calculation of the pinning potential for the DW, $U(X)$. This
is then used for the analysis of both the classical dynamics of
the DW and its semiclassical quantization. In
Sec.~\ref{s:continuum} the complementary continuum approach is
used to investigate the motion of the DW with finite velocity. The
DW linear momentum $P$ and the dispersion law of the DW, $E=E(P)$,
are calculated, and $E(P)$ is found to be a periodic function of
the momentum. The results for $U(X)$ and $E(P)$ are employed in
the next two sections to describe the dynamics of the DW in the
framework of the method of collective variables, assuming that the
DW is an effective quasiparticle with kinetic energy $E(P)$,
moving in the potential $U(X)$: In Sec.~\ref{s:free} the specifics
of forced DW motion as well as the detailed features of the band
spectrum of the DW in the case of the weak potential $U(X)$ are
discussed. The peculiarities of the dynamics of the DW in a finite
potential associated with the periodic character of $E(P)$ are
considered in Sec.~\ref{s:forced}. Here it becomes clear that
finite motion is typical for states which are either close to the
minimum or close to the maximum of the potential $U(X)$. In
Sec.~\ref{s:tunnel} we consider the quantum tunneling transitions
between these states, corresponding to an adjacent unit cell. In
this section the general character of the DW dispersion law
including the effects of lattice pinning is discussed.
Sec.~\ref{s:CR} gives our concluding remarks.

\section{The discrete model and a domain wall structure.}\label{s:discrete}

We start from the Hamiltonian describing a spin chain with Ising-type exchange
interaction in a magnetic field $\vec H$ directed perpendicularly to the easy
axis of a magnet. Spins $\vec S_n$ are located at points $na$ of a chain
(distance $a$, $n$ integer),
\begin{equation}
\label{ham}\widehat{H}=-J\sum_nS_n^zS_{n+1}^z-g\mu_B H\sum_nS_n^x.
\end{equation}
$J$ is the exchange integral, $g$ the gyromagnetic ratio and $\mu
_B$ the Bohr magneton. The dimensionless field $h = g\mu _B H/2JS$
will be used in the following (the notation $2 h = \gamma$ was
used in Ref.~\cite{mik-miyash})  This model is usually called
transverse field Ising model.

For classical spins the ground state of the model at $h<1$ ($g\mu_B H<2JS$) is
doubly degenerated with
\begin{equation}
\label{theta0} S_x=S\sin \theta_0, \ S_z=\pm S\cos \theta_0, \
\sin \theta_0 = h.
\end{equation}

We begin to describe the semiclassical motion of the DW within the
discrete spin model by studying the solutions for moving DW's in the
discrete classical model. We notice that the exchange part for the
model of Eq.~(\ref{ham}) is the simplest particular case of the
so-called XYZ-model with three independent parameters $J_1\neq J_2\neq
J_3$
\[%
\widehat{H}_{XYZ}=-\sum
(J_1S_n^xS_{n+1}^x+J_2S_n^yS_{n+1}^y+J_3S_n^zS_{n+1}^z).
\]%
For an isotropic FM we have $J_1=J_2=J_3$ and DW's strictly are absent
(however, 'pulse solitons' may exist) For the XXZ-model $(J_1=J_2 <
J_3)$ in the absence of the magnetic field the exact static solution
for the discrete model with classical spins has been constructed by
Gochev\cite{Gochev}.  As far as we know this discrete problem has not
yet been solved for $H \neq 0$ or for the more general classical
XYZ-model even for $H=0$.  The solution found by Gochev for the
XXZ-model reads $ S_n^z= S \tanh[\kappa (n-n_0)]$, with $\kappa =
\ln[(J_3+\sqrt{J_3^2-J_1^2})/J_1]$, where $1/\kappa$ measures the
domain wall thickness. In the Ising limit the DW thickness goes to
zero, $1/\kappa \rightarrow 0$ at $J_1\rightarrow 0$. We emphasize
that the value of $n$ which describes the DW position in this solution
(taken as $n=n_0$ above) is an arbitrary ({\em not necessarily
integer}) number. Thus the model with $H=0$ has the nontrivial
property that the energy of the DW does not depend on its center
position for arbitrarily large anisotropy. It is not clear, whether
this property is valid for the same Hamiltonian in the quantum case.

For the Ising model we easily understand the above result since at
$H=0$ the DW is described by the following solution: $S_n^z =-S$ at
$n<n_0$, $S_n^z =+S$ at $n>n_0$, however $S_n^z$ at $n=n_0$ may have
arbitrary values. If $S_{n_0}^z = 0$ the DW is localized on the spin
with $n= n_0$; on the other hand, for $S_n^z =-S$ or $S_n^z = +S$ at
$n=n_0$ the DW localized in the center of the bond which connects the
spin at $n=n_0$ with spins at $n=n_0+1$ or $n=n_0-1$,
correspondingly. For $H=0$ the energies of DW's centered on the spin
(central spin (CS) DW) or on the bond (central bond (CB) DW)
coincide. For all intermediate cases $S_n^z \neq 0,\ \pm 1$ at $n=n_0$
it is natural to postulate that the DW is localized on a point $X$,
which does not coincide with a site or the bond center and to describe
the DW dynamics in terms of its coordinate $X$ treated as a collective
variable. Let us introduce $X$ in the following way: we choose some
lattice site and define the DW located on this site to have the
coordinate $X=0$. Let us then find the $z$-projection
$\mathcal{S}_{tot}^{z \ (0)}$ of the total spin $\mathcal{S}_{tot}^z$
of the DW localized on that site. It is then natural to determine the
coordinate of any DW via the value of the total $z$-projection of the
spin $\mathcal{S}_{tot}^z$, associated with this DW from the
expression $X=a(\mathcal{S}_{tot}^z-\mathcal{S}_{tot}^{z \ (0)})/2S$.
In order to account appropriately for possible divergencies in
infinite chains the difference
$\mathcal{S}_{tot}^z-\mathcal{S}_{tot}^{z \ (0)}$ is calculated as
$\sum_{n=-\infty}^{n=\infty} [S^z_n(X) -S_n^z(X=0) ]$ (here $S_n^z(X)$
and $S_n^z(X=0)$ are the $z-$ projection of spins on site $n$ for the
DW with the coordinate $X$ and the DW with $X=0$ resp.). Then, for the
DW in the model of Eq.~(\ref{ham}) with $H\neq 0$ the complete
definition reads
\begin{equation} \label{X-definition}
X = \frac{a}{2S\cos\theta_0}
\sum_{n=-\infty}^{n=\infty} [S^z_n(X) -S_n^z(X=0) ].
\end{equation}
For zero field the DW energy $E$ does not depend on the coordinate
$X$. At $H\neq 0$, as shown in Ref.~\onlinecite{mik-miyash}, a
dependence of the DW energy on its coordinate emerges, pointing to the
presence of a lattice pinning potential $U=U(X)$. The pinning
potential $U(X)$ has its minimum value for $X=na$, i.e. for the CS DW,
and its maximum value for $X=(a/2)(2n+1)$, i.e. for the CB
DW.\cite{mik-miyash} In order to investigate the nonstationary
dynamics of the DW we will treat $U(X)$ as a potential energy. The
analysis of the related kinetic energy will be discussed in the next
section.

When the DW is moving in a discrete chain it is natural to use the
periodic potential $U(X),\ U(X+a)=U(X)$. The form of the potential can
be chosen as in \cite{br-loss},
\begin{equation}
\label{u(x)}U(x)=U_0\sin ^2(\pi X/a),
\end{equation}
where $U_0$ characterizes the intensity of pinning caused by
discreteness. With the choice of Eq.~(\ref{u(x)}) for the
potential, the value of $U_0$ can be found as the energy
difference between the static central spin and central bond
DW's.\cite{mik-miyash} In order to calculate $E_{CS}$ and E$_{CB}$
we need in the solution of the corresponding classical discrete
problem, which is known for small magnetic field
only.\cite{mik-miyash} Moreover, for our purpose we need to know
not only the energy difference $E_{CS}-E_{CB}$, but the full
dependence $U(X)$ in order to verify the dependence of
Eq.~(\ref{u(x)}).

\begin{figure}[tb]
\includegraphics[width=70mm]{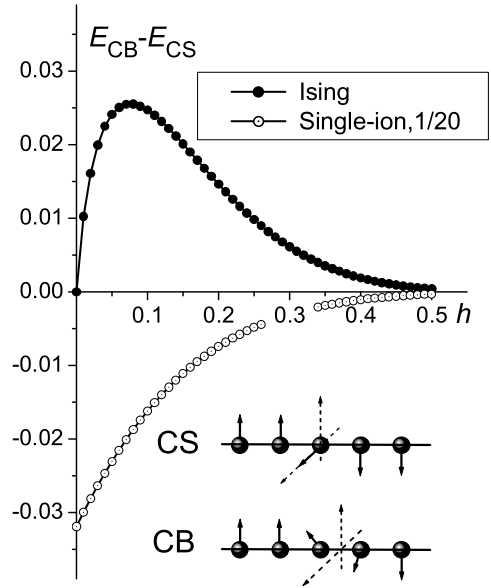}
\caption{\label{fig:Eob-os+} The difference of the energies (in units
of $JS^2$) for CB and CS domain walls (shown schematically at the
bottom of the figure) vs. magnetic field. The corresponding
dependence for the model with single-ion anisotropy, normalized by 20,
is also given for comparison (open circles).}
\end{figure}

To reveal $U(X)$ we have carried out a numerical analysis of the
model of Eq.~(\ref{ham}) in accordance with the definition
Eq.~(\ref{X-definition}) for the coordinate $X$. We have searched
the conditional minimum of the energy as given by Eq.~(\ref{ham})
for a finite spin chain at a fixed value of $\mathcal{S}_{tot}^z$.
To solve this problem, the simplex type method with non-linear
constraints was chosen as the method of minimization. This method
is based on the steepest descent routine applied to functions of a
large number of variables, and it is able to find the conditional
minimum for a given function with fixing a small number of
combinations to given values. The method exhibits a fast
convergence when a spin distribution with only one domain wall
placed near the point of inflection is used as starting condition.
For our problem, the angular coordinates of each spin were chosen
as variables, and the minimum of the energy was found with fixing
the value of $\mathcal{S}_{tot}^z$. The DW was created by fixing
the direction of two spins on the ends of the chain corresponding
to Eq.~(\ref{theta0}). The chain length $N$ varied from 20 to 100,
and for the the case of interest, $H \leq JS$, the result was
independent of the chain length for $N \geq 30$. The value of
$U_0$ in Eq.~(\ref{u(x)}) was determined from the difference
$E_{CS}-E_{CB}$; the behaviour of this quantity is shown in
Fig.~\ref{fig:Eob-os+}.

\begin{figure}[tb]
\includegraphics[width=70mm]{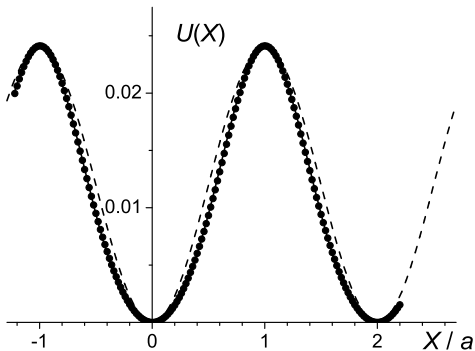}
\caption{\label{fig:U(X)} Shape of the domain wall pinning potential
$U(X)$ (in units of $JS^2$) for the dimensionless magnetic field
$h=0.05$. Symbols denote the numerical data. The empirical dependence
$0.024\sin^2(\pi x/2a)$, and a fit for a more general dependence with
$U_0 = 0.024,\ U_1 =- 0.00255$ are shown as dashed and full lines,
respectively.}
\end{figure}

The analysis shown in Fig.~\ref{fig:U(X)} demonstrates that in the
case of interest to us, $h \ll 1$, $U(X)$ is fairly well described by
Eq.~(\ref{u(x)}).  If we consider a more general dependence, allowing
for one more Fourier component,
\begin{equation}
\label{u(x)2}U(x)=U_0\sin ^2(\pi X/a)+U_1\sin ^2(2\pi X/a),
\end{equation}
the dependence of $U(X)$ is reproduced with a deviation of less than
$0.1\%$. However, as the correction related to $U_1$ is small, we will
use in the following mainly the simplest expression, Eq.~(\ref{u(x)}).

Note, as $U_0\rightarrow 0$ for $h\rightarrow 0$, and for
$h\rightarrow 1$ the transverse field Ising model is characterized by
a rather small pinning potential. For comparison, in
Fig.~\ref{fig:Eob-os+} the corresponding dependence is presented for a
ferromagnet with an isotropic Heisenberg interaction and a single ion
anisotropy of the form $W_a = -\sum_n K(S_n^z)^2$, which has the same
amplitude as Eq.~(\ref{u(x)}) at $K=J$. For this model $U_0$ has a
maximum at $H=0$; the typical values of $U_0$ are approximately 20
times higher than for the transverse field Ising model. Furthermore,
the quantity $U_0$ has opposite sign, i.e., in contrast to the
transverse field Ising model one has $E_{CB} < E_{CS}$.

\section{Domain wall dynamics in continuum approximation.}
\label{s:continuum}

Most of the results for the dynamics of domain walls or kink-type
solitons in magnets have been obtained in the continuum approach,
replacing $\vec S_n(t)$ by the smooth variable spin density $\vec
S(x,t)$, where $x$ is the coordinate along the chain. Within this
approach the dynamics of the vector field $\vec S(x,t)$ is described
by the well known Landau-Lifshitz equation without dissipation
\cite{LL} (see also \cite{phys.rep,sowsci})
\begin{equation}
\label{lan lifsh}
\frac{\partial \vec S}{\partial t}=\frac a\hbar
\left( \vec S\times \frac{\delta W\{\vec S\}}{\delta \vec S} \right).
\end{equation}
Here $W\{\vec S(x,t)\}$ is the ferromagnetic energy as functional of
the spin density. For our model, the transverse field Ising chain,
$W\{\vec S(x,t)\}$ corresponds to the Hamiltonian of Eq.~(\ref{ham})
and is written as: \cite{mik-miyash}
\begin{equation}
\label{energy} W\{\vec S\}=\int \frac{dx}a\left\{
J\frac{a^2}2\left( \frac{
\partial S_z}{\partial x}\right)^2-JS_z^2-g\mu _B HS_x\right\}.
\end{equation}

Using the continuum approach allows us to find a solution, which
describes a DW moving with a given velocity $V$. On the other hand,
the discreteness effects are evidently lost going from Eq.~(\ref{ham})
to Eq.~(\ref{energy}).

We use the relation $\vec S^2=S^2=\text{const}$ to write $\vec S=S
\: \vec m(x,t)$ i.e. to express the spin through the unit vector
field $\vec m(x,t)$. Its direction is determined by two
independent variables, we will use the angular variables $\theta $
and $\varphi$,
$$
m_z=\cos \theta, \;
m_x=\sin \theta \cos \varphi,  \; m_y=\sin \theta \sin \varphi.
$$
In terms of these variables the Landau-Lifshitz Eq.~(\ref{lan lifsh})
reads
\begin{eqnarray}
\label{angular}
\frac{\hbar S}a\sin \theta \frac{\partial \theta }{\partial t}
&=&\frac{\delta W}{\delta \varphi }, \nonumber  \\
\frac{\hbar S}a\sin \theta \frac{\partial \varphi }{\partial t}
&=&-\frac{\delta W}{\delta \theta },
\end{eqnarray}
where $W\{\theta ,\varphi \}$ is the ferromagnetic energy written as a
functional of $ \theta $ and $ \varphi $. These equations can be
considered as classical Hamiltonian equations for the canonically
conjugate variables $\cos \theta $ (momentum) and $\varphi$
(coordinate), with Hamilton function $W$.

Kink solitons in such a continuum approach have been described for a
number of magnetic chain models. These include the Heisenberg chain
with two single ion anisotropies
\cite{Sklyanin79,Mikeska80,EtrichMi83}, the Ising chain with
transverse exchange breaking the $xy-$symmetry \cite{ElstnerMi89} and
the $xy-$like Heisenberg chain with an external symmetry breaking
field.  \cite{EtrichMi88,EtrichMiMaThWe85} The qualitative results for
kink solitons with permanent shape are analogous in all these models:
The dispersion relation, energy vs. velocity, has two branches and one
distinguishes between the lower energy domain wall (LDW) and the upper
energy domain wall (UDW). The energies of these two solutions merge at
the maximum possible velocity $V = V_c$. A more natural formulation
results when the momentum is introduced instead of the velocity: then
the dispersion relation becomes single valued and periodic with the
magnetic unit cell.\cite{Haldane86} In particular the dispersion
relations can be given explicitly for the Heisenberg chain with two
single ion anisotropies. \cite{Sklyanin79,Mikeska80}

Eqs.~(\ref{lan lifsh}) or (\ref{angular}) are usually considered
as purely classical equations. To discuss their applicability to
the quantum regime one may use the quantum-field approach based on
the spin coherent states formalism.\cite{fradkin} Then the quantum
phase (Berry phase) appears, leading to qualitative effects such
as the suppression of quantum fluctuations for antiferromagnetic
chains \cite{affleck} and small magnetic particles with
half-odd-integer total spin.\cite{quant-interf} In this approach
the spin state on every site $n$ is determined by the spin
coherent state $|\vec m \rangle$, for which $\vec S \, |\vec m
\rangle = S \,\vec m \,|\vec m \rangle$. Here $\vec m$, as before,
is a unit vector, $\vec m^2=1$. In this approach, the dynamics of
the mean value of spin $\vec S=S\vec m$ is described by a
Lagrangian which can be written in the form
\begin{equation}
\label{lagr:S}
\mathcal{L}\{\vec S\}= \hbar S\int \frac{dx}a
              \vec {A}\dot {\vec S} - W\{\vec S \} ,
\end{equation}
where
\begin{equation}
\label{lagr:A}
\vec{A}(\vec S) = \frac{\vec{n}\times\vec{S}}{S (\vec S \vec{n}+ S)}.
\end{equation}
$\vec n$ is a unit vector with arbitrary direction, denoting the
quantization axis for coherent states. It is important that $\vec A$
has the form of the vector potential of a magnetic monopole field in
the full space $\{\vec S\}$ (not subject to the constraint $\vec
S^2=\text{const}$).

The vector potential has a singularity (Dirac string) for $\vec{S}\,
\vec{n} = -S$, i.e., on a half-line in space $\{\vec S\}$. Usually,
the "north pole" gauge with $\bm{n} = \bm{e}_z$ is used, then the
quantity $\vec{A}(\vec S) \, \dot{\vec S}$ acquires the familiar form
\begin{equation}
\label{lagr:angular}
\vec {A}\dot {\vec S} = (1-\cos \theta )(\partial \varphi /\partial t).
\end{equation}

In the saddle point approximation for the Lagrangians
Eq.~(\ref{lagr:S}) or (\ref{lagr:angular}) one recovers the
classical Landau-Lifshitz equations in the form of Eq.~(\ref{lan
lifsh}) or (\ref{angular}) for the mean value of spin $\vec S$.
The potential $\vec{A}$ of the monopole field permits gauge
transformations (in particular, the change of the direction of the
spin quantization axis $\vec n$ and, hence, the positions of
singularities). These do not change the equations of motion, but
make a contribution to the Lagrangian in the form of a total time
derivative of some function of spin. This may in principle be
significant for the calculation of the DW momentum.  Naturally,
the classical equations are not affected by the gauge transform.
On the other hand, the term with $\partial \varphi /\partial t$ in
Eq.~(\ref{lagr:angular}) is of crucial importance for the quantum
dynamics of domain walls in spin chains with half-odd integer
spins.\cite{br-loss} For example, this term is responsible for the
destructive interference of paths for DW tunneling from one
minimum of the crystal potential to an adjacent one.\cite{br-loss}

In order to construct a solution corresponding to a DW moving with
velocity $V$, we have to write down the Landau-Lifshitz equations in
angular variables, and to restrict ourselves to traveling wave-type
solutions, $\theta =\theta (\xi ),\ \varphi =\varphi (\xi),\ \xi
=x-Vt$, with the natural boundary conditions, $S^x\rightarrow S\sin
\theta_0$ at $\xi \rightarrow \pm \infty$, and
$$
S^z \rightarrow \sigma S\cos \theta_0,\
S^z \rightarrow -\sigma S\cos \theta_0
$$
at $\xi \rightarrow \infty$ or $\xi \rightarrow -\infty$,
respectively.  Here $\sigma=\pm 1$ is $\pi_0$, the topological charge
of the DW and $\cos \theta_0$ characterizes the ground state, see
Eq.~(\ref{theta0}). Then, using Eq.~(\ref{energy}), the equations for
$\partial \theta /\partial t=-V \theta ^{\prime}$ acquire the simple
form $ v \, \theta^{\prime}=-2h\sin\varphi $. Here and below the prime
denotes differentiation with respect to $\xi $, and we have introduced
the dimensionless variable
$$
v=V/V_0, \ V_0= aJS/\hbar
$$
and put $a=1$ in all intermediate equations. We will restore the
dimensional parameters in final results only.

The set of Eqs.~(\ref{angular}) with the traveling wave ansatz has
one integral of motion, which after taking boundary conditions into
account can be written as
\begin{equation}
\label{IntMotion}
\frac{1}{2} [(\cos \theta)^{\prime }]^2 - \sin^2 \theta
            + 2h \sin \theta \cos \varphi =  h^2 = \text{const}.
\end{equation}
Using the simple relation between $\theta^{\prime}$ and $\sin \varphi $, this
can be rewritten in the variable $\theta $ only. Finally we arrive at the
simple differential equation for $\theta $,
\begin{equation}
\label{solution}
(\cos \theta )^{\prime}=\sigma _1 \sqrt{2\sin^2 \theta-v^2}
                +\sigma_2 \sqrt{2h^2-v^2}.
\end{equation}

$\sigma_{1,\ 2} =\pm 1$ are two independent discrete parameters which
determine the topological charge of the kink as $\pi _0$-topological
soliton and the spin direction in the kink center, $\xi =0$, with
respect to the magnetic field. Thus $\sigma_{1,\ 2}$ fixes the type of
the DW (i.e. DW with lower energy and DW with upper energy, see next
paragraph).

Eq.~(\ref{solution}) determines, in particular, the maximum possible
value of the velocity of the DW ($\pi $-kink), the critical velocity
$v_c$
\begin{equation}
\label{v c}v_c=\sqrt{2} h
\end{equation}

Eq.~(\ref{solution}) can be integrated in terms of elementary
functions, generalizing the solution for $v=0$ as given before.
\cite{PrelovsekS81,mik-miyash} The analysis shows that the equation
has both LDW and UDW solutions as described above. For $v\neq 0$ the
solutions are obtained by substituting
\begin{equation}
\label{GenSol1}\cos \theta =\cos \psi \cdot\sqrt{1-v^2/2},\  \sin
\alpha =\sigma _2\sqrt{\frac{v_c^2-v^2}{2-v^2}}\ ,
\end{equation}
and can be given in the implicit form
\begin{equation}
\label{GenSol2} \sqrt{2}\sigma _1(\xi -\xi _0)=\psi +\tan \alpha
\cdot \ln \left| \frac{\sin [\left( \psi -\alpha \right) /2]}{\cos
[\left( \psi +\alpha \right) /2]}\right|
\end{equation}

As for $v=0$ \cite{mik-miyash} and in related models (see above) there
are two types of DW's, with different energies. For $v=0$, the DW with
lower energy (LDW) in its center has the direction of spin $\vec S(0)$
parallel to the magnetic field $\vec H$, whereas for the DW with
higher energy (UDW) $\vec S(0)$ is antiparallel to $\vec H$. This
applies similarly for moving DW's, however, $\vec S(0)$ and $\vec H$
are not exactly parallel resp.  antiparallel for $V\neq 0$.\\

As for $v=0$, the solution for the upper DW has a discontinuity in the
space derivative $\theta^{\prime}$.  At the critical velocity $v=v_c$
LDW and UDW become identical and the solution $\theta(\xi)$ can be
given in explicit form:
$$
\cos\theta=\sigma \cos\theta_0\cdot\sin(\sqrt{2}\xi/a) \quad
\text{at} \; \vert \xi \vert <a\pi /2\sqrt{2},
$$
and
$$\cos\theta=\sigma \cos\theta_0\cdot (\xi/|\xi |) \quad \text{otherwise}.$$
It has a discontinuity in the second space derivatives.

\begin{figure}[tb]
\includegraphics[width=70mm]{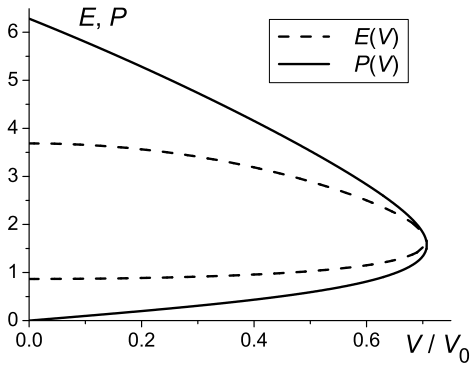}
\caption{\label{fig:E(V)} The dependence of the DW energy (in
units $JS^2$) and its momentum (in units $\hbar S/a$) on its
velocity for the value of $h=0.5$. }
\end{figure}

In order to find energy and momentum, the DW parameters of interest,
we do not need the solution for LDW and UDW in explicit form. It is
easy to use Eq.~(\ref{solution}) and to pass from the integrals over
$\xi$ to the integration over $\theta $. Then an elementary
calculation of the kink energies gives
\begin{eqnarray}
\label{E_low}
E_{lower}&=&JS^2\sqrt{2}\left[ \left(
1-\frac{v^2}2\right) \arcsin
\sqrt{\frac{1-h^2}{1-v^2/2}}\right]- \nonumber \\
&-&JS^2\sqrt{\left( 1-h^2\right) \left( 2h^2-v^2\right) }\ , \\
\label{E_up}
E_{upper} &=& E_{lower}+4hJS^2\sqrt{\left(
2-v^2\right)\left( 1-\frac{v^2}{v_c^2}\right) }\ ,
\end{eqnarray}
for the lower and upper DWs resp.. Thus, the dependence $E(v)$
consists of two branches, $E_{lower}$ and $E_{upper}$ which merge at
$v=v_c$, and the full dependence $E=E(v)$ is a continuous
double-valued function, see Fig.~\ref{fig:E(V)}. For the most
interesting case of small magnetic fields, $h\ll 1$, or $v\leq v_c \ll
v_0 $, it is given by the unified equation
\begin{equation}
\label{E-smallH}
E(V)=\sqrt{2}JS^2\left[ \arcsin \sqrt{1-h^2}
              \mp 2h\sqrt{1-\frac{v^2}{v_c^2}}\right],
\end{equation}
where the sign $\mp $ corresponds to the lower and upper branches of
$E=E(P)$ respectively.

In the three-dimensional ferromagnet the UDW is unstable. However,
this instability develops as an inhomogeneous perturbation (in the
plane of the DW) and is not present for DW's in one-dimensional
magnets. Below we will show that for the more natural representation
of the DW energy, namely as a function of its momentum, the function
$E(P)$ is single-valued, and the upper branch just corresponds to the
larger values of momentum. This explains its stability in the
one-dimensional case.

The kink momentum is determined as the total field momentum of the
magnetization field,\cite{Haldane86}, i.e.
\begin{equation}
\label{P_general}
P=- \int d\xi \; \frac{\delta \mathcal{L}} {\delta \dot {\vec S}}
      \; \vec S^{\prime}= -\int \frac{1}a \, \vec {A} \, d\vec S.
\end{equation}

The dynamical part of the Lagrangian in Eq.~(\ref{lagr:S}) and the
expression (\ref{P_general}) for the momentum display
singularities associated with the singular behavior of the vector
potential $\vec A$. The vector potential for a monopole inevitably
has a singularity on a line (Dirac string) and moreover is not
invariant with respect to gauge transformations. The kink momentum
also seems not to be invariant under these gauge transformations.
If one uses a Lagrangian written in angular variables, problems
due to the non-differentiability of the azimuthal angle $\varphi $
at the points $\theta = 0$ and $\theta = \pi $ appear. (This
problem is also important for the theory of moving two-dimensional
topological solitons, see Ref.~\onlinecite{papaniko}). It is
important, however, that the difference of the momenta for two
different states of the DW is a gauge-invariant
quantity.\cite{GalkinaIv00} To show this, let us imagine the DW as
a trajectory in spin space $\{\vec S\}$, $\vec S=\vec S(\xi)$. The
trajectories emerging from one point, say, $\vec S=S(\vec e_x\sin
\theta _0 -\vec e_z\cos \theta _0$) and ending at another point
$\vec S=S(\vec e_x\sin \theta _0 +\vec e_z\cos \theta _0$) can be
associated with DW's that move with different velocities but obey
identical boundary conditions at infinity. In this case, the kink
momenta are determined by integrals of the form $\int \vec A \,
d\vec S$ along these trajectories. It is clear that the difference
of the momenta is determined by the integral $\oint \vec A \,
d\vec S$ along a closed contour. According to Stokes' integral
theorem, the integral in question can be represented as the flux
of the vector $\vec B = \; \text{curl}\vec A$ through the surface
bounded by this contour, $\int \vec B \; d{\vec \Omega }$. Here
the integral is taken over that region on a sphere which is
bounded by the trajectories corresponding to the two kinks in
question. The vector $\vec B = \; \text{curl} \vec A = \vec
S/|\vec S|^3$ involves no singularities. Returning to angular
variables, this difference can be written as $(\hbar S/a)\cdot
\int \sin \theta \, d\theta \, d\varphi$, that is just the area on
the sphere. Thus, the dependence $V(P)$ or $E(P)$ has been
reconstructed, apart from the arbitrariness to choose the
reference point for the momenta.

\begin{figure}[tb]
\includegraphics[width=70mm]{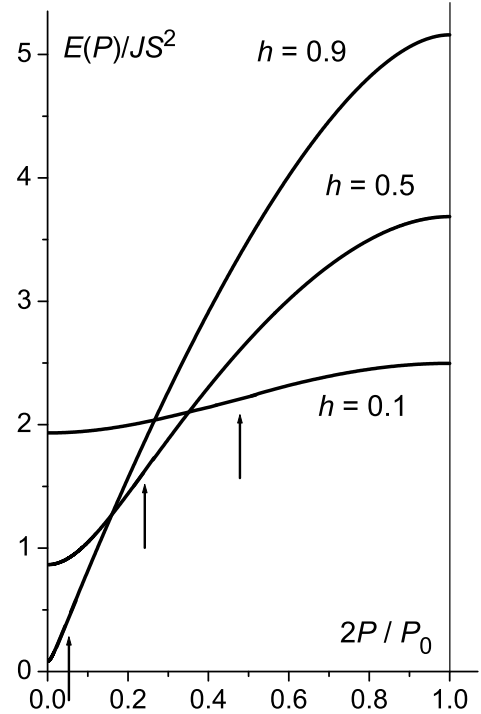}
\caption{\label{fig:E(P)free} The dependence of the DW energy on
its momentum for different values of the magnetic field (given
next to the curves); arrows indicate the position of the
momentum corresponding to the critical velocity $v_c$.}
\end{figure}

The trajectories describing $v=0$ kinks appear to belong to the large
circle passing through the poles $\theta =0,\ \pi$ and the end point
of the magnetic field vector $\vec H$. Obviously, the difference of
the momenta $P$ for the immobile LDW and UDW is determined by the area
of the half-sphere, $\Delta P = 2\pi \hbar S/a$. All the remaining
trajectories that correspond to moving kinks cover the regions between
these trajectories. In particular, kinks at the critical velocities
$\pm v_c$ correspond to two non-planar trajectories. The $v = \pm v_c$
trajectories can be reached by proceeding from either type of $v = 0$
domain walls, LDW or UDW. Thus, we can see that $E(v)$ has two
branches, one characterized by a higher energy and the other
characterized by a lower energy. These two branches merge at the
critical velocity $v = v_c$. If we assume $P = 0$ for the $v = 0$ LDW,
then the momentum grows up to $P = \pm P_c, \; P_c < 2\pi\hbar S/a$ as
the absolute value of the kink velocity increases to $v_c$. As we
proceed further along the upper branch of the dependence $E(V)$, the
kink velocity decreases, while the momentum increases further to $P =
\pm 2\pi\hbar S/a$ when the velocity approaches again zero. When we
continue, we begin to cover the area of the sphere once more and the
momentum $P$ grows, while the kink energy takes the same values as
before: thus we do indeed arrive at a periodic dependence $E(P)$ with
period $P_0$ determined by the total area of the sphere. The value of
$P_0$ depends only on the spin value $S$ and the lattice spacing $a$,
\begin{equation}
\label{p-zero}
P_0=\frac{4\pi \hbar S}a.
\end{equation}
The presence of two planar solutions (with $\varphi = \text{const}$)
describing the DW's with $v = 0$ and different energies is a common
feature for any model with uniaxial anisotropy subject to a transverse
magnetic field. As we know, for Heisenberg exchange interaction and
single-ion anisotropy, only numerical solutions describing mobile DW`s
can be found.\cite{IvKulagin} But even in this case the periodicity
continues to be the same. On the other hand, for the Ising model
considered here, the exact solution is known, and the momentum can be
calculated explicitly. As a result the momentum for LDW and UDW takes
the following form
\begin{eqnarray}
\label{P_low}
P_{lower}&=&\frac{2\hbar S}a \arcsin \left( \frac
{v}{v_c}\sqrt\frac{1-h^2}{1-v^2/2}\right) -\nonumber \\
&-&\frac{v\hbar S\sqrt 2}{a}\arcsin \sqrt {\frac{1-h^2}{1-v^2/2}}
\end{eqnarray}
\begin{equation}
\label{P_up}
P_{upper}=P_{lower}+\frac{4\hbar
S}a\sqrt{1-\frac{v^2}2}\cdot \arcsin \sqrt{1-\frac{v^2}{v_c^2}}.
\end{equation}
These equations together with Eqs.~(\ref{E_low}, \ref{E_up}) give us
the dispersion law for the DW in implicit form as shown in
Fig.~\ref{fig:E(P)free}.

For the mostly interesting case of small magnetic fields $h\ll 1$, or
$v \leq v_c \ll v_0 $ the dependence $P(v)$ can be given as the
unified expression
\begin{equation}
\label{P-smallH}
P_{l,\ u}=\frac{2\hbar S}a\left( \frac \pi 2\pm
\arcsin \sqrt{1-\frac{v^2}{v_c^2}}\right),
\end{equation}
where $\pm$ correspond to UDW and LDW, respectively.  Finally,
limiting ourselves to the case $h\ll 1$, the dispersion law for both
DW's takes the form
\begin{equation}
\label{e(p)}
E(P)=\sqrt{2}JS^2\left[ \frac \pi 2- 2h\cos \left(
\frac{2\pi P}{P_0}\right) \right]
\end{equation}

\begin{figure}[tb]
\includegraphics[width=70mm]{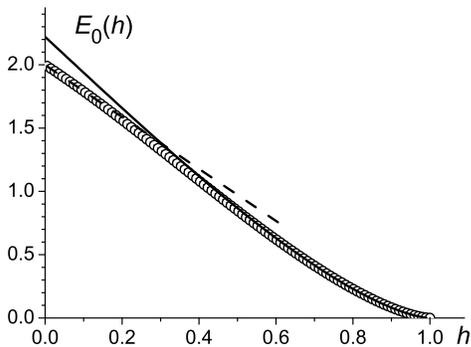}
\caption{\label{fig:E(H)} The dependence of the DW energy with zero
velocity (in units $JS^2$) on the dimensionless magnetic field $h$;
symbols depict the numerical data for the discrete model; the solid
line is the theoretical result in the continuum approximation
Eq.~(\ref{e(p)}); the approximation of numerical data by the linear
function $E=2(1-h)JS^2$ is shown as dashed line.}
\end{figure}

This equation will be used in the remaining part of the paper to
describe the classical dynamics of domain walls and to perform
their semiclassical quantization. It is based on the continuum
approximation and this is the only conceivable approach to
analytically describe moving DW's; however, the validity of the
continuum approximation has to be justified before doing so since
it seems to be non-adequate for the Ising model ferromagnet. This
applies in particular to the limit $h\ll 1$, when the DW thickness
becomes comparable or even less than the lattice spacing $a$. To
check the applicability of the continuum approximation we present
in Fig.~\ref{fig:E(H)} a comparison of the DW energies as obtained
from the numerical approach to those obtained from
Eq.~(\ref{e(p)}) at $P=0$. The discrepancy is seen to be of the
order of $10\%$ for $h \rightarrow 0$ and thus surprisingly small.
In addition, the linear dependence of the DW energy on the
magnetic field, $E(H)=JS^2(A \pm hB)$ is valid for both
approaches, with values $A \approx 2.0$ and $B \approx 2.0 $ for
the numerical data and $A = \pi/\sqrt 2 \approx 2.2$ and $B =
2\sqrt 2 \approx 2.8 $ from Eq.~(\ref{e(p)}). As we will see
below, the use of the dependence $E(P)=JS^2[A-hB\cos(2\pi
P/P_0)]$, with "improved" coefficients $A$ and $B$, gives much
better agreement with the numerical data for the dispersion law of
quantum solitons.\cite{mik-miyash}

\begin{figure}[tb]
\includegraphics[width=70mm]{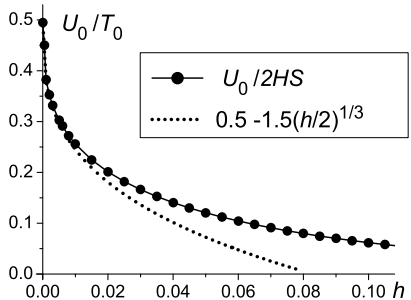}
\caption{\label{fig:ToUo} The ratio of the maximum value of the
pinning potential to the quantity $T_0$ characterizing the dispersion
of the DW. For $T_0$, the "improved" value of the coefficient $B=2$ is
used. Also shown as dotted line is the limit of this quantity for
extremely small values of the magnetic fields \cite{mik-miyash},
$U_0/T_0=0.5[1-3(h/2)^{1/3}]$.}
\end{figure}

One more important parameter, the maximum value of the pinning
potential $U_0$ was treated numerically for arbitrary values of
$h$. For extremely small values of $h$,
$U_0=E_{CB}-E_{CS}=2JS^2h[1-3(h/2)^{1/3}]$ was obtained from the
analytical expressions for $E_{CS}$ and $E_{CB}$ (see Eqs.~(11,12)
in Ref.~\onlinecite{mik-miyash}). As we will see below, the ratio
of $U_0$ to the difference of the DW energies with $P=0$ and
$P=P_0/2$, $T_0=JS^2hB$ is an important parameter for the
description of the DW dynamics. Whereas the dependence of $U_0$ on
$h$ is non-linear at small values of $h$ because of the presence
of the non-analytical term proportional to $h^{1/3}$, the DW
"kinetic energy" $T_0$, is linear in $h$, see Sec.~\ref{s:forced}
below. The maximum value of $U_0/T_0=1/2$ is realized in the limit
$h \rightarrow 0$, but in fact this important ratio decreases fast
to rather small values when $h$ grows (see Fig.~\ref{fig:ToUo}).

In the limit of small values of momentum, $P\rightarrow 0$, the
parabolic approximation can be used and the energy takes the form
$E=P^2/2M$, where the effective mass of a kink is \begin{equation}
\label{mass} M=\frac{4\sqrt 2 \hbar ^2}{a^2hJ}.  \end{equation} It is
seen from Eq.~(\ref{mass}) that the effective mass turns to infinity at
$h \rightarrow 0$. This is one more manifestation that in a purely
uniaxial model of a ferromagnet in the absence of a transverse
magnetic field domain wall motion is impossible.\cite{IvKulagin} The
use of the parabolic approximation (\ref{mass}) seems to be adequate
for DW's  moving with small velocities. However, some important features
are lost in this approximation: For example, the correct value of the
energy bands forming the DW spectrum should be found within the analysis
of the full periodic dependence $E(P)$.

We note that for small $h$ the value of the effective mass for the
upper DW branch, defined by $E_{upper}(P)=E(P_0/2)-(P-P_0)^2/2M_{up}$
agrees with $M$ from Eq.~(\ref{mass}). However, $M$ and $M_{up}$ are
substantially different at finite $h$, see Fig.~\ref{fig:E(P)free}
above.

To conclude this section, we emphasize that within the macroscopic
classical approach the dispersion law (\ref{e(p)}) exhibits a
periodic dependence of $E(P)$ which, from Bloch's theorem, is
characteristic for discrete quantum models. Moreover the maximum
value of momentum, $P_0=4\pi \hbar S/a$, which may be called the
size of magnetic Brillouin zone, and the size on the crystalline
Brillouin zone $P_B=2\pi \hbar /a$ have values of the same order
at $S \sim 1$. The expression for $P_0$ contains Planck's constant
$\hbar $ and the lattice constant $a$ and formally seems to
characterize both the quantum nature and the discreteness of the
model. As discussed above, this is not the case: if one accounts
for the {\em macroscopic} (continuous) character of the
magnetization $M_0=g\mu _BS/a$ (otherwise the discussion of
ferromagnetism is meaningless), $P_0$ can be written using only
$M_0$ and the classical gyromagnetic ratio $\gamma =e/2mc,
P_0=4\pi M_0/g\gamma $.

\section{Forced motion and dispersion relation of domain walls }\label{s:free}

The periodic dependence discussed in the preceding section means that
one can restrict the values of momentum $P$ to the magnetic Brillouin
zone, $-P_0/2<P<P_0/2$. On the other hand, as with Bloch electrons, in
order to analyze the motion under the influence of an external force,
it is useful to consider that the DW momentum obeys the equation
$dP/dt=F_e$, where $F_e$ is external force, and to allow that $P$
increases without limits beyond the first Brillouin zone. The
expression for the DW energy $E(P)$ can be used to describe the DW
dynamics in the spin chain in terms of the DW coordinate $X$
considered as a collective variable. Its dynamics is governed by the
Hamiltonian
\begin{equation} \label{dw-hamiltonian}
\mathcal{H}(P,X)=E_0(h)+T_0\sin^2 (\frac{\pi P}{P_0}) + U(X),
\end{equation}
where $E_0(h)$ is the DW energy at $P=0$. $E_0(h)$ has no effect on
the dynamics of the DW coordinate $X$ and will be omitted below. Here
we have introduced the parameter
$$ T_0=4\sqrt{2}hJS^2=2\sqrt{2} g\mu_0 SH $$
which describes the magnitude of free DW dispersion, and added the
potential $U(X)$, without specifying its physical origin. For the
discrete spin chain, the potential is the pinning potential
originating from the discreteness effects introduced above.

It is easily seen that the relation $\dot X \equiv V=\partial
\mathcal{H}/\partial P$ as in analytical mechanics immediately gives
the periodic (oscillating) dependence of the DW velocity on DW
momentum
\begin{equation}
\label{V(P)} V=\frac{\pi T_0}{P_0}\sin (\frac{2\pi P}{P_0})
\end{equation}
Inverting this equation we recover the expression for the momentum
(\ref {P-smallH}) for small magnetic field.

The dynamical equation for the Hamiltonian reads $dP/dt=-\partial
\mathcal{H}/\partial X$. Choosing different forms of the
potential $U=U(X)$ one can consider different problems such as the
interaction between a DW and the inhomogeneities in the medium or with
external magnetic field $H_z$ directed along the easy axis. In the
last case we have
\begin{equation}
\label{force}
U(X)= - 2g\mu _BSH_z \cdot X/a, \ \text{and} \
\dot P=2g\mu _BH_zS/a.
\end{equation}

Therefore the DW velocity under the action of {\em constant}
magnetic field (constant force) oscillates with time. Note that
for a DW in the uniaxial ferromagnet this equation is nothing but
the one-dimensional version of familiar Slonczewsky
equations.\cite{malslon} It describes the nontrivial properties of
DW dynamics, such as the oscillating motion of DW's as
response to a constant external force. These effects were observed
in a number of experiments on DW dynamics in bubble-materials, see
Ref.~\onlinecite{malslon}. Formally, such a motion corresponds to
the Bloch oscillations well known for quantum mechanical
electron in an ideal crystal lattice. Bloch oscillations for
solitons in different media were recently reviewed by
Kosevich.\cite{Kosevich01} Also, such effects for DW's in spin
chains with $S=1/2$ have been discussed from the viewpoint of
Bloch particles recently.\cite{KirLoss98} However, it is important
to note that for DW's in a continuum model the origin of the
effects is different in principle: It is not associated with
discreteness and it exists even in the continuum limit. Thus,
even in the classical continuum model the dynamics of a DW
exhibits a number of properties peculiar to Bloch particles
(electrons), i.e. to quantum objects moving in the periodic
potential of the crystal lattice. On the other hand, the time
dependence of this forced DW velocity is oscillating with the
classical Larmor frequency $\Omega_L=g\gamma H_z$. Combining Eqs.
(\ref{V(P)}) and (\ref{force}) one can find
$$ V(t)=V_{max}\sin\Omega_Lt, \ V_{max}=\pi T_0/P_0.$$
Thus, the quantum and classical regularities intertwine in a very
intricate manner in the problem of domain wall dynamics.

To analyze the quantum DW motion in the spin chain including
discreteness effects, we will use Eq.~(\ref{dw-hamiltonian}) as a
quantum Hamiltonian, using the periodic potential of Eq.~(\ref{u(x)})
and $U_0$ as found in the Sec. \ref{s:discrete}.  Let us start with the
case of nearly free motion, when $U_0\ll T_0$. In this case the DW
energy is given not by the usual momentum, but instead by the
quasi-momentum $P$ which is determined only up to a reciprocal lattice
vector. According to Bloch's theorem, the energy $E(P)$ should be
periodic with period $P_B=2\pi \hbar /a$.  When $U_0\rightarrow 0$,
the dispersion law is described by Eq.~(\ref{e(p)}), or
Eq.~(\ref{dw-hamiltonian}) with $U_0=0$, for almost all $P$ values
(with small corrections $\sim U_0^2$). Only when the non-perturbed
dispersion laws intersect, $E(P)=E(P+nP_B)$ with $n$ integer,
corrections become essential $(\sim U_0)$ and energy gaps appear.

We now want to discuss the situation for different values of spin $S$,
since the ratio $P_0/P_B = 2S$ depends on the spin value. Strictly
speaking, small spin values, e.g. $S=1/2,\ 1,...$, cannot be
described in the frame of our semiclassical approach. But we argue
that our approach will be valid at least qualitatively also for these
cases.

For spin $S=1/2$ we have $P_0=P_B$, and there is no intersection of
the non-perturbed dispersion law with its image shifted by $P_B$ (the
intersection of two parabolas in Ref.~\onlinecite{br-loss} is an
artefact of the parabolic approximation). In this case the effect of
small $U(X)$ on the DW dispersion relation $E(P)$ is only quadratic in
the small parameter $U_0$ and is negligibly small. For all other spin
values, $S>1/2$, we have $P_0>P_B$ and dispersion curves $E_0(P)$
extended periodically with period $P_B=2\pi \hbar /a$ do
intersect. However, all intersections occur on the boundaries of the
Brillouin zone, namely, at $P=\pm \pi \hbar /a$. It is clear that in
this case the DW spectrum contains $N_b=2S$ energy bands.

\section{Classical dynamics of domain walls in a finite potential
and its quantization}\label{s:forced}

In this section we will analyze the classical DW dynamics and its
quantization for an arbitrary (finite) periodic potential of the form
given in Eq.~(\ref{u(x)}). The most important ingredient to this
analysis is that the Hamiltonian of Eq.~(\ref{dw-hamiltonian}), just
as the classical energy, is a periodic function of momentum, and that
there is an upper bound for the energy. As already mentioned, this
property is already present in the classical theory of DW motion and
has nothing to do with quantum mechanics. We will show below that this
property leads to results which are qualitatively different from those
obtained in the standard quadratic approximation, both for the pure
classical DW motion as well as for its quantization.

Let us consider the dynamical system described by the classical
Hamiltonian of Eq.~(\ref{dw-hamiltonian}), taking $U(X)$ into
account. The corresponding Hamilton equations
$$
\frac{\partial P}{\partial t}=-\frac{\partial \mathcal{H}}{\partial X},
\;
\frac{\partial X }{\partial t}=\frac{\partial \mathcal{H}}{\partial P},
$$
have an obvious integral of motion,
\begin{equation}
\label{hamiltonian2}
\mathcal{H}(P,X)=T_0\sin^2 \frac{\pi P}{P_0} +
             U_0\sin^2 \frac{\pi X}{a}= \text{const}.
\end{equation}
Here we have omitted the constant part $E_0(h)$ which does not affect
the equations of motion. These equations cannot be integrated in terms
of elementary functions, however, a sufficiently complete
understanding of the DW dynamics can be found using the phase plane
analysis. This system is characterized by the periodicity in momentum
$P$ and by the presence of an upper bound for the Hamiltonian. In view
of this its dynamics shows characteristic features which do not
manifest themselves for standard dynamic systems with a parabolic
dependence on momentum.

\begin{figure}
\includegraphics[width=70mm]{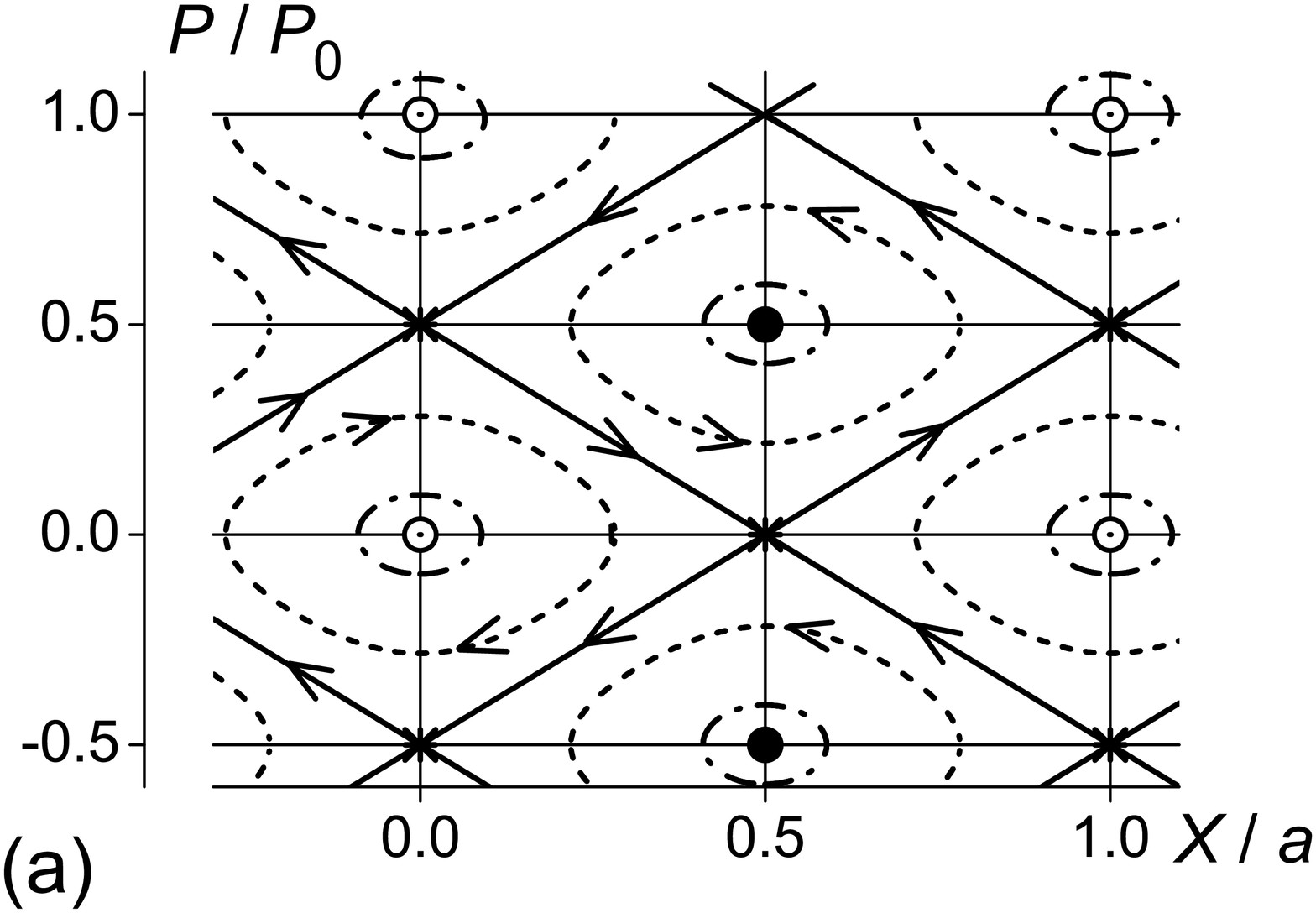}
\includegraphics[width=70mm]{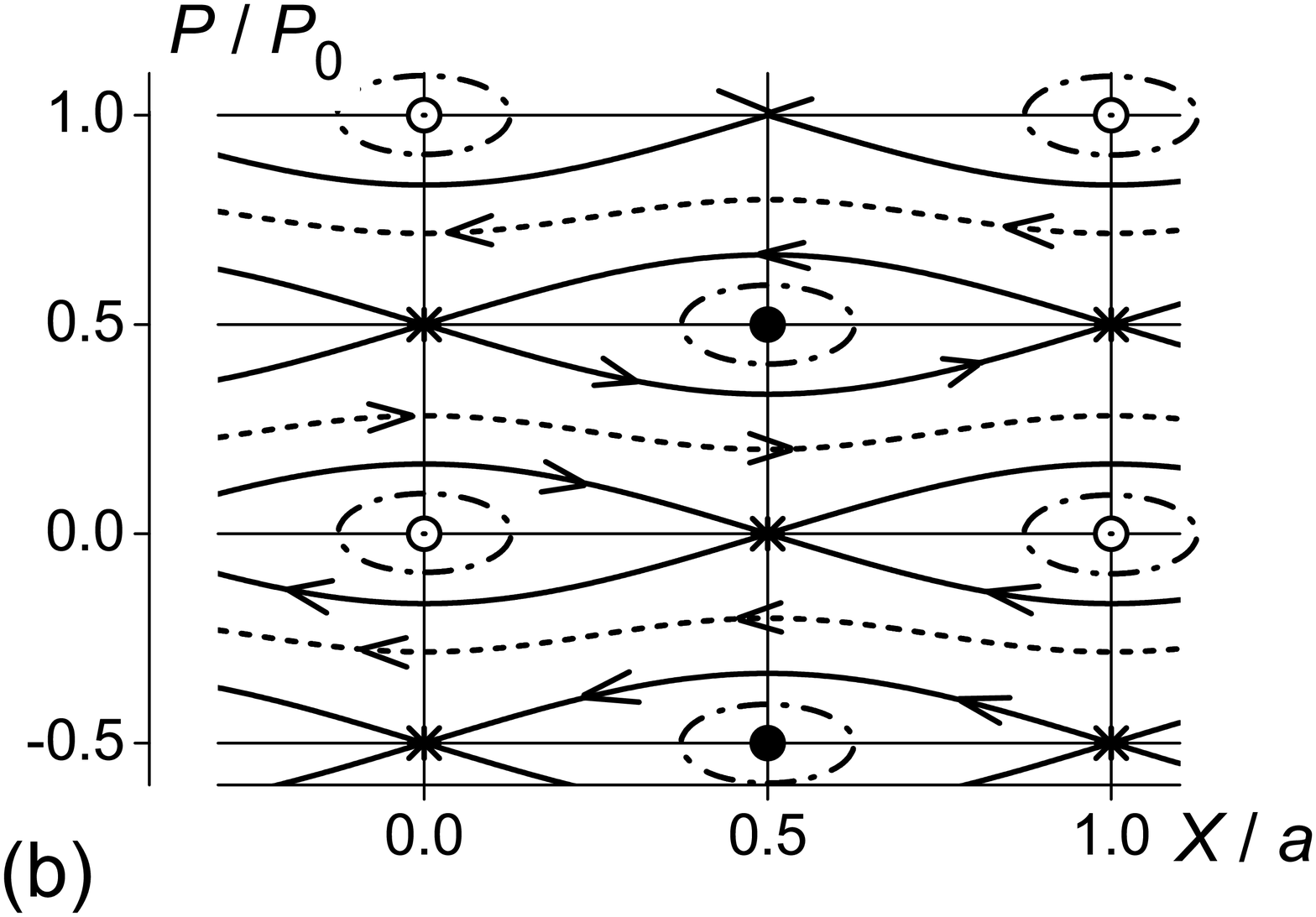}
\caption{\label{PhasePlane}
The phase plane $(P,X)$ for the DW dynamics for (a) $U_0=T_0$ and (b)
$U_0=0.25T_0$. Separatrix trajectories are shown by full lines, other
trajectories by dashed lines; full and open circles denote the
positions of center-like singular points $C1$ and $C2$, crosses
correspond to the saddle points.}
\end{figure}

It is easy to show that at arbitrary $U_0/T_0$ there are two sets of
center-like singular points in the phase plane, see
Fig.~\ref{PhasePlane}. One of them ($C1-$type center) corresponds to
the minimum of the potential $U(X)$ and the minimum of the "kinetic
energy" $T(P)=T_0\sin^2 (\pi P/P_0)$,
$$ C1: X=an,\ P=P_0m,\ \mathcal{H}_{C1}=0, \hfill $$
the other set of centers ($C2-$type center) is located at the maximum
of both potential and kinetic energies,
$$ C2:\ X=a(2n+1)/2,\ P=P_0(2m+1)/2,\ \mathcal{H}_{C2}=U_0+T_0.$$
Here and below $n,m$ are integers. The existence of $C2-$type centers
describing the steady small oscillations of the UDW near the potential
{\it maximum}, is a unique property of Hamiltonian systems with an
upper bound in the Hamilton function $\mathcal{H}(P,X)$.

There are also two sets of saddle points in the phase plane, $S1$
and $S2$. Points $S1$ correspond to the  maxima of the potential
$U(X)$ and the minima of $T(P)$
$$  S1;\ X=a(2n+1)/2,\ P=P_0m,\ \mathcal{H}_{S1}=U_0, $$
points $S2$ correspond to the minima of $U(X)$ and the maxima of
$T(P)$,
$$  S2;\ X=an,\ P=P_0(2m+1)/2,\ \mathcal{H}_{S2}=T_0. $$

At the allowed values of the integral of motion,
$\mathcal{H}_{C1}\leq \mathcal{H}\leq \mathcal{H}_{C2}$, the phase
plane $(P,X)$ separates into regions with different types of
motion. Two types of finite motion are present within one
interatomic distance (one period of the potential). One type of
finite motion corresponds to the oscillations of the LDW near the
minimum of the potential, it requires $\mathcal{H}_{C1}\leq
\mathcal{H}\leq \mathcal{H}_{S1}$ and is standard in the
analytical dynamics of a particle. (We suggested $U_0 \leq T_0$
here).  A second type of finite motion, namely oscillations of the
UDW near the potential maximum, is realized for $
\mathcal{H}_{S2}\leq \mathcal{H}\leq \mathcal{H}_{C2}$. These
regimes are separated from the rest of the phase plane with
different types of infinite motion by separatrix trajectories,
ending in one of the saddle points. The only exception is the case
$U_0=T_0$ or $\mathcal{H}_{S1}=\mathcal{H}_{S2}$ when the
separatrix trajectories connect saddle points of different type
and form a two-dimensional net; as a consequence infinite motion
is absent. For the transverse field Ising model with small $U_0$
this is hard to realize, see Sec.~\ref{s:discrete} (in particular
Fig.~\ref{fig:ToUo}), but for methodological purposes we find it
convenient to start with this case.

In this case we have $T_0=U_0$, and one easily verifies that the area
of the phase plane (mechanical action) per unit cell and one period
of momentum is equal to $ \mathcal{A}_0=4\pi \hbar S$. Thus, from the
Bohr-Sommerfeld quantization rules, there are ($\mathcal{A}_0/2\pi
\hbar )=2S$ quantum eigenstates of a DW, related to its oscillations
per one unit cell. It is clear that if one takes into account
tunneling transitions between equivalent points of the lattice,
these localized states will turn into energy bands.

This result coincides with the result obtained in the previous section
for the opposite limit of an infinitely weak pinning potential. An
additional argument on the number of energy bands is obtained from the
well-known exact result for so-called Harper equation,\cite{harper}
familiar in the problem of electronic quantum motion in a periodic
potential in the presence of a finite magnetic field. In accordance
with Ref.~\onlinecite{harper} the problem is reduced to a Hamiltonian
as in Eq.~(\ref{hamiltonian2}), ${\mathcal H}_H=\sin ^2(p/2)+\sin
^2(\pi \beta q)$. For the rational case of the Harper equation, at
$\beta =m/n$, the eigenvalue spectrum for the Hamiltonian ${\mathcal
H}_H$ shows $n$ non-overlapping bands. The simple canonical
transformation $p\rightarrow \pi P/P_0,\ q\rightarrow XP_0/\pi $,
leads to our Hamiltonian (\ref{hamiltonian2}) with $\beta =1/2S$. This
gives immediately the above result, $N_b=2S$.

For $T_0<U_0$, a case more realistic for the Ising model, the
situation is different: in addition to the localized trajectories in
phase space, trajectories corresponding to the infinite motion above
the potential barrier appear, see Fig.~\ref{PhasePlane}b. These
trajectories describe the overbarrier dynamics of DW's in the pinning
potential. It is clear, that such states should be described well by
the model of the nearly-free particle, discussed in the previous
section.

The other limiting case, large $U_0$, i.e. $U_0>T_0$ is hard to
realize in the transverse field Ising model. On the other hand, it
could be interesting for other models of ferromagnets supporting DW
states, and we want to discuss it briefly. The phase plane for large
$U_0/T_0$ can be obtained from Fig.~\ref{PhasePlane}b by replacing
$P/P_0$ by $x/a$ and vise versa. Then, the topology of the phase plane
is fundamentally changed: trajectories with finite changes of the kink
coordinate and {\em infinite} range in momentum, appear. For these
trajectories the DW coordinate oscillates near certain positions,
which do not coincide with a minimum or maximum of the pinning
potential. This is nothing but Bloch oscillations in the pinning
potential $U(X)$.

At this point we want to emphasize that the semiclassical result
$\mathcal{A}_0=4\pi \hbar S$ does not depend on the details of the
model. Only the periodicities of the Hamiltonian in $P$ and $X$ with
periods $P_0$ and $a$ are essential for the argument. The
considerations presented in the previous section are model independent
as well, and lead to the same value of $P_0$ for all transverse field
models. Therefore the final result for the number of energy bands
remains valid for more general models of ferromagnets subject to a
transverse magnetic field. Thus the number $N_b$ of energy bands for
DW's in all these models should be equal to $2S$. This is in agreement
with the numerical result for the kink dispersion law in the quantum
transverse field Ising model with spin $S=5$, where 10 energy bands
were obtained. \cite{mik-miyash}

\section{DW tunneling}\label{s:tunnel}

In the classical case, there are states corresponding to finite
motion (oscillations) of DW's near the extrema of the potential in
the phase plane both at small and large values of $U_0$. This
applies to the LDW near the minimum of $U(X)$ and as well to the
UDW near the maximum of $U(X)$. Owing to the symmetry present in
the Hamiltonian (\ref{hamiltonian2}), $P\rightarrow P_0/2-P$,
$X\rightarrow a/2-X$, their dynamics is described similarly;
therefore one can consider only, say, the LDW case. In the quantum
case, the tunneling of DW's from one site to another becomes
possible, and these states form the energy bands. It is clear that
for extremely small $U_0$ the nearly free approximation discussed
in section \ref{s:free} is valid. But the standard situation for
semiclassical systems (like DW's for large spin ferromagnets) is
the tight-binding limit, for which the probability of tunneling is
small, and the states with energy $E_n$ are almost localized in a
potential well. In order to estimate $E_n$, we consider the
parabolic approximation for both $U(X)$ and $T(P)$. In this case
we have $E_n=\hbar \omega_0\cdot n$, where
$$
\omega _0=\sqrt{\frac{1}{M}
\left(\frac{d^2U}{dx^2}\right)_{X=0}}=\frac{\sqrt{U_0T_0}}{4\hbar S}.
$$
$M$ is the effective mass of Eq.~(\ref{mass}). Both $U_0$ and $T_0$
are proportional to $S^2$, therefore $ \omega _0 \propto S$ and for
semiclassical spins $S\gg 1$ the value of $E_n$ for $n \ll S$ can be
smaller than $U_0$ even in the case $U_0\ll T_0$. Then the width of
the $n^{\rm th}$ energy band, $\Delta E_n$, resulting from the
tunneling between the quantized energy levels $E_n$, is smaller that
the value of $T_0$ and even $E_n$. It is clear that this first of all
should correspond to the lowest level $E_n$, i.e. $n=0$ (the "ground
state of the DW" in the pinning potential), but it could be true as
well for some higher levels with $n>1$.

In order to understand the possibility of semiclassical dynamics of
the lower DW, consider the limit of small $U_0$, $(U_0 \ll T_0)$.  The
area $\mathcal{A}_L$ under the separatrix trajectories $S1$ is easily
found and the number of quantum states for the finite motion of the
LDW, $N_L=\mathcal{A}_L/2\pi \hbar $ is defined by the following
expression
$$
N_L=\frac{8S}{\pi ^2}\sqrt{\frac{U_0}{T_0}}.
$$
This is of the same order of magnitude as the value obtained from
$N_L\cdot\hbar \omega_0 = U_0$. The same expression is obtained for
the number of states of the upper DW, localized near the potential
maximum, $ N_U=N_L$. At $U_0 \ll T_0$, the values of $N_U$ and $N_L$
are much smaller than the total number of DW states $2S$, but can
still be large compared to 1: When the inequality
$$
\frac 1{S^2}\ll \frac{U_0}{T_0} \ll 1,
$$
(meaningful for $S\gg 1$) holds, the states of finite motion are a
small fraction of all DW states, but $N_U=N_L \gg 1$. It is easy to
estimate the mean values of $P^2$ and $X^2$ in these states,
$$
\left\langle P^2\right\rangle =
P_0^2\cdot \frac 1{2\pi S}\sqrt{\frac{U_0}{T_0}}, \; \;
\left\langle X^2\right\rangle =a^2\cdot \frac 1{2\pi S}\sqrt{\frac{T_0}{U_0}}.
$$

In view of that, it is not easy to satisfy the condition of
applicability of the parabolic approximation for both $U(X)$ and
$T(P)$, $\left\langle P^2\right\rangle \ll (P_0^2)$ and
$\left\langle X^2\right\rangle \ll a^2$ simultaneously. Thus, in
order to investigate the tunneling splitting of the levels it is
preferable to use the full non-parabolic Hamiltonian
Eq.~(\ref{hamiltonian2}). The best way to analyze such models is
the instanton technique, which can be applied for arbitrary field
theories, see Ref.~\onlinecite{Rajaraman}. In this approach, the
amplitude of probability $\mathcal P_{12}$ of the transition of
the DW from one given state to another is determined by the path
integral $ \int DX\cdot \exp\{i{\mathcal A}[X]/\hbar \}$, where
${\mathcal A}[X]=\int dt\cdot {\mathcal L}[X(t)]$ is the
mechanical action functional. Here ${\mathcal L}(X,\ \dot X)$ is
the Lagrangian describing the dynamics of the DW coordinate $X$,
and the integration $DX$ goes over all configurations $X(t)$
satisfying the given boundary conditions. The Lagrangian
${\mathcal L}(X,\ \dot X)$ is easily obtained, as in standard
mechanics, from ${\mathcal L}= P \dot X-{\mathcal H}$: express $P$
and ${\mathcal H}$ in terms of $\dot X$, using the known
expressions for $P=P(V)$ and ${\mathcal H}=E(V)+U(X)$ (see section
\ref{s:continuum}), replacing $V$ by $\dot X$.

It is convenient to make a Wick rotation $t\to i\tau $, passing to the
Euclidean space-time.  Then one has the Euclidean propagator $\int
DX\,\exp [-{\mathcal A}_E/\hbar ]$, where the Euclidean action is
written as
$$
{\mathcal A}_E=\int {\mathcal L}_E(X,\Omega )d\tau ,
$$
with $\Omega =dX/d\tau$. For the LDW this gives
\begin{equation}
\label{lagrEu}
{\mathcal L}_E=\frac{2\hbar S}{a} \left[\Omega \text{arcsinh}
\left(\frac \Omega{V_c}\right) +V_c-\sqrt{V^2_c+\Omega^2}\right]+U(X).
\end{equation}
Constructing the instanton solution can then be viewed as minimization
of ${\mathcal A}_E$ with respect to $X(\tau)$. The minimum of the
Euclidean action is realized on the solution of the Euler-Lagrange
equation for ${\mathcal L}_E$. The first integral of this equation has
the form
$$
\frac{2\hbar S}{a}\left(\sqrt{V_c^2+\Omega^2}-V_c \right)-U(X)={\mathcal C},
$$
with ${\mathcal C}=0$ on the instanton solution. Then, the expression
for ${\mathcal A_E}$ reduces to the integral
$$
{\mathcal A}_E =
   \frac{2\hbar S}{a}\int_0^a \text{arcsinh}(\frac{\Omega }{V_c})dX,
$$
which has to be calculated using the connection between $\Omega$
and $X$ from the first integral, $(\Omega
T_0/V_c)^2=4U(X)[T_0+U(X)]$. Finally, the expression for the
Euclidean action on the instanton solution takes the form
\begin{equation}
\label{ae final}
\frac{{\mathcal A}_E}{\hbar}
   =\frac{8S\sqrt{U_0}}{\pi}\cdot \int_0^{\pi /2} \frac{\psi d\psi
   \sin \psi}{\sqrt{T_0+U_0\sin^2\psi}}.
\end{equation}

\begin{figure}[tb]
\includegraphics[width=70mm]{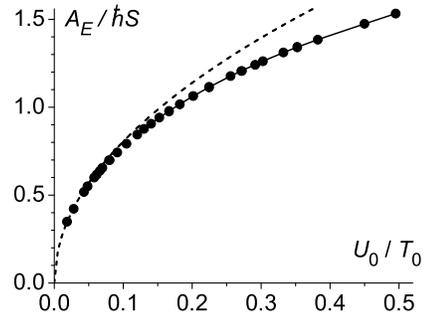}
\caption{\label{fig:A_E} The dependence of the Euclidean action on
the value  of $U_0/T_0$ as found numerically (symbols) and from the
asymptotic Eq.~(\ref{ae parabolic}), $A_E \propto \sqrt{U_0/T_0}$,
valid for small $U_0/T_0$ (dotted line). The maximum value of
$A_E/ \hbar S = 1.533$ is reached at $U_0/T_0=0.5$, that is the
maximum value of $U_0/T_0$ for the transverse field Ising model.}
\end{figure}

For the most natural case $U_0\ll T_0$ this gives the result
\begin{equation}
\label{ae parabolic} {\mathcal A}_E^{(0)}=\frac{8\hbar S }{\pi
}\cdot \sqrt{\frac{U_0}{T_0}} \
 \text{at}\ U_0\ll T_0,
\end{equation}
as usual for a Lagrangian with quadratic dependence on the velocity.
For arbitrary $U_0/T_0$ the calculations are more complicated but the
tendency is seen as follows: ${\mathcal A}_E$ is smaller than predicted
by Eq.~(\ref{ae parabolic}) for the same value of $U_0/T_0$, see
Fig.~\ref{fig:A_E}. Thus, the band width for finite $U_0/T_0$ is
larger than follows from Eq.~(\ref{ae parabolic}).

For energy bands $\Delta E_n$ with $1 \ll n \ll N_L$ and $U_0\ll T_0$
we can use the fact that in the regime of classical motion the simple
parabolic dispersion law is valid to calculate the tunneling splitting
of the levels. This allows to use the standard expression for the
tunnel splitting of the levels (see Ref.~\onlinecite{quantum mech},
problem 3 after \S 50) and results in
$$ \Delta E = (\hbar \omega _0/\pi )
   \exp \left\{-\int_{-a^{\prime}}^{+a^{\prime }}\{2m[U(X)-E]\}^{1/2}
                 dX/\hbar \right\}.
$$
Here $a^{\prime }$ is the turnover point of the classical trajectory
with energy $E_n$ (corresponding to a non-split level) defined by the
equation $U(a^{\prime })=E$. It is convenient to rewrite it using the
Euclidean action only. After some simple algebra we arrive at the
following universal formula:
\begin{equation}
\label{semiclass}
\Delta E_n \simeq \hbar \omega _0 \left(
 \frac{{\mathcal A}_E}{\hbar } \right)^{n+1/2} \cdot
\exp \left( -\frac{{\mathcal A}_E}{\hbar }\right) \ ,
\end{equation}
where ${\mathcal A_E}$ is given by Eq.~(\ref{ae final}) or
Eq.~(\ref{ae parabolic}), with ${\mathcal A_E} \propto S$ in both
cases. This equation shows that the energy splitting for $n>1$
continues to be exponentially small in $S$, but with larger
pre-exponential factor. This is in good agreement with numerical
data.\cite{mik-miyash}

\section{Summary and discussion.}\label{s:CR}

\begin{figure}[tb]
\includegraphics[width=70mm]{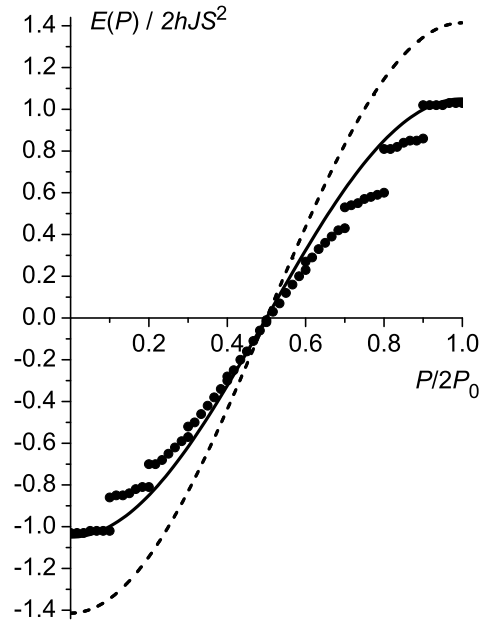}
\caption{\label{fig:E(P)free1} The dependence of the DW energy on its
momentum for the magnetic field $h=0.05$ as obtained in the continuum
approximation. The dashed line presents the result of the continuum
approach, the solid line is improved by using the fitted dependence
$E(h)$ for small $h$ (see Fig. \ref{fig:E(H)}). Symbols show the
numerical data for the dispersion law of the quantum
DW.\cite{mik-miyash}}
\end{figure}

We start this summary with the explicit results obtained for the
transverse field Ising model, and then go on to discuss the
conclusions which might be obtained for more general models. The
characteristic feature of the Ising model is that the two
important parameters $U_0$ and $T_0$ of the theory are governed by
only one quantity, the dimensionless magnetic field $h$;
practically always $U_0\ll T_0$ holds. However, the behavior of
the quantum DW spectrum depends also on the value of the atomic
spin $S$. If $U_0 \ll T_0$, but $ S^2 > T_0/U_0$, top and bottom
energy bands are very narrow. The neighboring bands are narrow
also, but $S$ timer wider, and the intermediate part of the
spectrum consists of wide bands with narrow energy gaps. This
picture corresponds well with the numerical data for the model
with $S=5$ of Ref.~\onlinecite{mik-miyash} (see
Fig.~\ref{fig:E(P)free1}).

Unfortunately, the direct quantum mechanical approach used in
Ref.~\onlinecite{mik-miyash} can be applied for small values of $h$
only; then the discreteness effects are large and the continuum
approximation is not appropriate. But the discrepancies are not too
large, and even in this unfavorable situation we can claim at least
qualitative agreement. The data are fitted quantitatively only with
use of fitted dependence $E(h)$ (see end of Sec.~\ref{s:discrete} and
Fig.~\ref{fig:ToUo}). It is worth to mention that there are at least
two reasons for discrepancies, caused by (i) nonapplicability of the
continuum limit to the small field discrete classical Ising model, and
(ii) the fact that the Landau-Lifshitz equation is not adequate to
describe quantum DW's.  We believe that discreteness is the main
source of the discrepancy, and for models with wide enough DW's, (such
as models of ferromagnets with nearly-isotropic exchange interaction
and weak magnetic anisotropy), the description of quantum dynamics of
DW's based on the scheme proposed in this paper will be at least
semi-quantitative.

At this point it is interesting to compare the results of the
instanton approach with data from an analysis of the model for
extremely small field, when spin zero-point fluctuations far from a DW
can be neglected and the one-site approximation can be
applied.\cite{mik-miyash} In both approaches the widths of the energy
bands with lower energies are exponentially small in the spin
magnitude $S$. The one-site approximation yields an exponential factor
of the form of $\exp(-2S \ln 2) \approx
\exp(-1.39S)$,\cite{mik-miyash} which is quite close to the value
$\exp (-\mathcal {A}_E/\hbar) \rightarrow \exp(-1.53S)$ at $h
\rightarrow 0$, obtained in the instanton approach using improved data
for $T_0$, $U_0/T_0 \rightarrow 1/2$. It is interesting to note that
using the data obtained directly from the continuum approximation,
$U_0/T_0=1/(2 \sqrt 2 )$, the factor takes the form $\exp(-1.34S)$
which is much closer to the value obtained in
Ref.~\onlinecite{mik-miyash}. The agreement between the data can be
considered rather good, regarding the conceptual difference between
the two approaches.

Good agreement is also found with respect to the fact that the band width
increases with energy, the ratio of the widths of neighbouring energy
bands is given by a factor of $S$. But the pre-exponential factors for
these, essentially alternative, approaches differ by a factor of
$\sqrt S$. It is most likely that taking into account fluctuations far
from the DW center becomes essential for the determination of the
pre-exponential factor. It is worth noting the formal agreement
that, within the instanton approach, this factor is determined by the
fluctuation determinant only, again, far from the instanton
center. Again, the compliance of data of these two approaches can be
considered as satisfactory.

The instanton analysis demonstrates that while the field is
increasing, $\mathcal {A}_E/\hbar$ very rapidly decreases, see
Fig.~\ref{fig:A_E}. Thus, the width of the energy band grows rapidly,
the gaps become narrow and the nearly-free limit can be realized for
all energies. At $S=5$ the value of $\mathcal {A}_E/\hbar$ becomes of
the order of unity and $\Delta E_0 \approx \hbar \omega _0$ already at
$U_0/T_0 < 0.008$ (equivalent to the magnetic field becoming close to
$h \sim 0.3$). But for any values of the parameter $h$, $2S$ energy
bands are present in the DW spectrum. As we argued above, this result
is not specific to the Ising model. It is mostly due to the value of
the period $P_0$, which is universal for all models of uniaxial
ferromagnets in a transverse magnetic field. Thus, the result that
2$S$ energy bands exist should be valid for all transverse field
models of ferromagnets.

The results obtained here for ferromagnets can be applied to Ising
antiferromagnets as well. The simple canonical transformation $$
S_n^{y,z}\rightarrow (-1)^n \sigma_n^{y,z},\ S_n^{x}\rightarrow
\sigma_n^{x}, $$ results in the same commutation relations for
staggered magnetization $\vec \sigma_n$ as for usual spins $\vec
S_n$ and reproduces the same form for the Hamiltonian,
Eq.~(\ref{ham}), replacing $\vec S_n$ by $\vec \sigma_n$ and
changing the sign of $J$. It is easy to prove, after the
appropriate gauge transformation, that the form of the Lagrangian,
Eq.~(\ref{lagr:A}) is also unchanged. Thus, all the results
obtained here for the DW dynamics, its dispersion law and its
semiclassical quantization are valid for Ising antiferromagnets as
well. Note that such a simple connection can be true only for
models with a Hamiltonian depending on two components of spin
only. This excludes near Heisenberg antiferromagnets. For the near
Heisenberg antiferromagnets with weak anisotropy of any origin,
exchange or single-ion, treated on the basis of the sigma-model,
the singularities in the Lagrangian for the staggered
magnetization are absent for a sufficiently general set of models
(including antiferromagnets subject to strong external magnetic
fields and in the presence of Dzyaloshinski-Moriya interactions of
arbitrary symmetry \cite{IvKireev2002}) and a periodic DW
dispersion law is not realized.\cite{GalkinaIv00}

Another interesting feature, which is common to classical and quantum
physics, is that states with finite range of motion for the
coordinate, but infinite range for momentum are present for large
enough pinning potential, $U_0>T_0$. Evidently such trajectories do
not exist for standard mechanical systems with a parabolic dispersion
law $ \mathcal{H}=P^2/2M+U(X)$. Of course the number of quantized
states of DW motion per unit cell does not depend on the ratio of
$U_0$ and $T_0$ and remains equal to $2S$. For the case of extremely
weak dispersion $ T_0\ll U_0$, it follows that intermediate DW
positions in the periodic potential are quantized and are equal to
$2S$ discrete values. Thus, in the case of semiclassical quantization
of DW motion within the continuum problem, there are $2S$ positions of
the DW, i.e. some internal scale, $a_m=a/2S$ appears in the
problem. This fact might be used to explain intuitively that a
magnetic Brillouin zone, $P_0=2SP_B$, appears (here $P_B=2\pi \hbar
S/a$ is the size of the Brillouin zone in a chain with interatomic
distance $a$). To explain this in a slightly different way, we mention
that the DW coordinate is directly connected with the $z-$projection
of the total spin $\mathcal{S}_{z,\ tot}$, see (\ref{X-definition}),
and this quantum-mechanical quantity can change by unit steps only.

It is interesting to compare the features established here for
transverse field models to the 180-degree DW ($\pi-$kink) in
ferromagnets with rhombic anisotropy as considered by Braun and Loss
\cite{br-loss}. For their model there are two different DW's with
lowest energy (Bloch DWs) and the corresponding $\vec S(\xi)$
trajectories are the halves of the large circle passing through the
easy axis and the axis with the intermediate value of anisotropy.

It is easy to argue, that in this case the dependence $E(P)$ found
in the continuum approximation is again periodic, but the period
of this dependence $P_{0, \pi}$ is determined by {\em
  half} of the sphere $\vec S^2=S^2$, and it is two times smaller than
that for the transverse field model, $P_{0, \pi} = 2\pi \hbar
S/a$.\cite{Sklyanin79,Haldane86,GalkinaIv00} The specific model
treated in \cite{br-loss} is exactly integrable, and the free
dispersion law $E(P)$ with this period can be constructed explicitly,
see Refs.~\onlinecite{sowsci,phys.rep}. On the other hand, for the
full classification of DW's, i.e. to describe the presence of two
different, but energetically equivalent DW's one can in this situation
introduce the discrete parameter {\em chirality}. Braun and Loss have
shown from a different argument that the minima of the dispersion laws
$E(P)$ for DW's with different chiralities have to be located at
different points of the $P$-axis, with distance equal to $P_{0, \pi}$
(in our terminology). Owing to the parabolic approximation, used in
Ref.~\onlinecite{br-loss}, the dispersion curves $E(P)$ for different
chiralities seem to be intersecting in some point. But accounting for
the real periodic dependence for $E(P)$, these two minima are smoothly
connected, with the maximum at the energy of the N\'{e}el DW, and no
intersection occurs. Thus, there is no reason for the chirality
tunneling to take place in this formulation. The deep connection of
chirality and linear momentum and the suppression of quantum tunneling
of chirality for free 180-degrees DWs in ferromagnets has been
discussed recently by Shibata and Takagi.\cite{TakagiShib00} It is
worth to note here, that also for 180-degree DW's in Heisenberg
antiferromagnets two non-equivalent DW's with equal energies are
present, but the dependence $E(P)$ is not periodic. Chirality is not
connected with the value of linear momentum and the effects of quantum
tunneling of chirality are present even without the pinning
potential.\cite{ivkol}

Finally we want to discuss the perturbation of the dispersion law for
a 180-degree DW caused by a small periodic pinning potential
$U(X)$. For this case, one has to locate the intersections of
non-perturbed spectra $E(P)$, periodically extended with period
$P_B=2\pi \hbar/a$. The situation here is slightly more complicated
than for transverse field models: For small spins $S=1/2$ and $S=1$
one has $P_{0, \pi} \leq P_B$, intersections are absent and no gaps
are present (this is the same situation as for the transverse field
model with spin $S=1/2$). For integer spins $S=k,\ k>1$ intersections
occur and energy gaps appear at the intersection points. However, all
these gaps are located on the boundaries of the Brillouin zone. Then,
the number of energy bands is equal to $S$. But for sufficiently large
half-integer spins $S=(2k+1)/2, k \geq 1$, one has
$P_0=P_B(2n+1)/2$, and some intersections are inside the Brillouin zone
at $P=(2n+1) \pi \hbar /2a$, with integer $n$. This leads to the
appearance of some additional gaps, which could be associated with the
tunneling of chirality. In this case the number of gaps increases and
for half-integer spin $S$ the number of energy bands in the dispersion
relation for 180-degree DW's is equal to $2S$.

\acknowledgments  When analyzing solitons in discrete model we
used the original software package created by A.~Yu. Merkulov
whose help is gratefully acknowledged. We wish to thank L.~A.
Pastur for helpful comments about the Harper equations and to A.~K
Kolezhuk for useful discussions. This work is supported in part by
the grant I/75895 from the Volkswagen-Stiftung.


\begin{references}

\bibitem{mikeska} H.-J. Mikeska and M. Steiner, Adv. Phys. {\bf 40}, 191
(1991).

\bibitem{sowsci} V.~G. Bar'yakhtar and B.~A. Ivanov, in {\em Soviet
Scientific Reviews, Section A. Physics}, I.~M.  Khalatnikov (ed.),
{\bf 16}, 3 (1993).

\bibitem{phys.rep}  A.~M. Kosevich, B.~A. Ivanov, and A.~S. Kovalev, Phys.
Rep. {\bf 194}, 117 (1990).

\bibitem{villain} J. Villain, Physica, {\bf B 79}, 1 (1975).

\bibitem{mikeska+} H.-J. Mikeska, S. Miyashita, and G. Ristow,
 J. Phys.: Condens. Matter, {\bf 3}, 2985 (1991)

\bibitem{bethe} H.~J. Bethe, Z. Phys. {\bf 71}, 205 (1931).

\bibitem{baxter}  R.~G.Baxter, Ann. Phys. {\bf 70}, 323, (1972);
J.~D. Johnson, S. Krinsky and B.~M. McCoy, Phys. Rev. {\bf A8}, 2526 (1973).

\bibitem{ivkol} B.~A. Ivanov and A.~K. Kolezhuk, JETP Lett. {\bf
60}, 805 (1994);  Phys. Rev. Lett. {\bf 74}, 1859 (1995); JETP
{\bf 83}, 1202 (1996); B.~A. Ivanov, A.~K. Kolezhuk, and V.~E.
Kireev, Phys. Rev. {\bf B 58}, 11514 (1998)

\bibitem{TakagiTat96} S. Takagi and G. Tatara,  Phys. Rev. {\bf B 54},
9920 (1996).

\bibitem{TakagiShib00} J. Shibata and S. Takagi, Phys. Rev. {\bf B 62},
5719 (2000).

\bibitem{Freire} J. A. Freire,  Phys. Rev. {\bf B 65} 104436 (1996).

\bibitem{br-loss}  H.-B. Braun and D. Loss, Phys. Rev. {\bf B 53}, 3237 (1996).

\bibitem{KirLoss98}  J. Kiriakidis and D. Loss, Phys. Rev. {\bf B 58},
5568 (1998).

\bibitem{mik-miyash}  H.-J. Mikeska and S. Miyashita, Z. Phys. {\bf B 101}, 275
(1996).

\bibitem{Gochev}  I.~G. Gochev, Sov. Phys.-JETP {\bf 58}, 115 (1983).

\bibitem{LL}  L.~D. Landau and E.~M. Lifshitz, Sow. Phys. {\bf 8}, 157
(1935).

\bibitem{Sklyanin79} E.K.~Sklyanin, LOMI preprint (1979).

\bibitem{Mikeska80} H.-J.~Mikeska in {\it Physics in One Dimension},
  eds. J.~Bernasconi and T.~Schneider, Berlin 1981

\bibitem{EtrichMi83} C.~Etrich and H.-J.~Mikeska, J.Phys. C: Solid State
  Phys. {\bf 16}, 4889 (1983).

\bibitem{ElstnerMi89} N.~Elstner and H.-J.~Mikeska, J.Phys.: Condens. Matter
  {\bf 1}, 1487 (1989).

\bibitem{EtrichMi88} C.~Etrich and H.-J.~Mikeska, J.Phys. C: Solid State
  Phys. {\bf 21}, 1583 (1988).

\bibitem{EtrichMiMaThWe85} C.~Etrich, H.-J.~Mikeska, E.~Magyari, H.~Thomas and
  R.~Weber, Z.Phys. B{\bf 62}, 97 (1985).

\bibitem{Haldane86} F.D.M.~Haldane, Phys. Rev. Lett. {\bf 57}, 1488 (1986).


\bibitem{fradkin}  E. Fradkin, {\em Field theories of condensed matter
systems}, in {\em Frontiers in Physics} vol. 82, Addison-Wesley,
Reading, MA, Chapter 5, 1991.

\bibitem{affleck}  F.~D.~M.Haldane, J. Appl. Phys. {\bf 57}, 3359 (1985); I.
Affleck, Nucl. Phys. {\bf B257}, 397 (1985); for a review see: I.
Affleck, J. Phys. Condens. Matter {\bf 1}, 3047 (1989); I.
Affleck, in: {\it Fields, Strings and Critical Phenomena} [ed. E.
Br\'ezin and J. Zinn-Justin, North-Holland, Amsterdam, 1990].

\bibitem{quant-interf}  D. Loss, D.~P. DiVincenzo, and G. Grinstein, Phys.
Rev. Lett. {\bf 69}, 3232 (1992); J. von Delft and C.~L. Henley,
Phys. Rev. Lett. {\bf 69}, 3236 (1992).

\bibitem{PrelovsekS81} P.~Prelovsek and I.~Sega, Physics Lett. {\bf 81A}, 407
  (1981).

\bibitem{papaniko} N. Papanicolaou and T.~N. Tomaras, Nucl. Phys. {\bf B 360},
425 (1991).

\bibitem{GalkinaIv00} E.~G. Galkina and B.~A. Ivanov, JETP Lett. {\bf 71}, 372 (2000).

\bibitem{IvKulagin} B.~A.Ivanov and N.~E. Kulagin, JETP {\bf 85}, 516 (1997).

\bibitem{malslon}  A.~P. Malozemoff and J.~C. Slonczewski, {\em Magnetic
Domain Walls in Bubble Materials}, Applied Solid State Science,
Supplement I, Academic Press, New York (1979).

\bibitem{Kosevich01} A.~M. Kosevich, Low Temp. Phys. {\bf 27}, 513
(2001).

\bibitem{harper}  M. Wilkinson, J.Phys.A.: Math.Gen. {\bf 27}, 8123 (1994).

\bibitem{Rajaraman}  R. Rajaraman, {\em Solitons and Instantons,\/}
(North-Holland, 1982).

\bibitem{quantum mech}  L.~D.Landau and E.~M.Lifshitz, {\em Quantum Mechanics},
Plenum Press (1994).

\bibitem{IvKireev2002}B.~A. Ivanov and V.~E. Kireev, JETP {\bf 94}, 270 (2002).

\end{references}
\end{document}